\documentclass[jcp,aip,amsmath,amssymb,reprint,USenglish]{revtex4-2}

\usepackage{graphicx}
\usepackage{floatflt,epsfig} 
\usepackage{mathtools}
\usepackage{color}
\usepackage{braket}
\usepackage{hyperref}
\usepackage[normalem]{ulem}
\usepackage[english]{babel}
\usepackage{blindtext}
\usepackage{lipsum}  
\usepackage{amsthm}
\usepackage{tabularx}
\usepackage{bm}
\usepackage{dsfont}
\usepackage{bbold}
\usepackage{ulem}
\usepackage{mwe,tikz}
\usepackage[percent]{overpic}
\usepackage{todonotes}
\usepackage{algorithm}
\usepackage{algpseudocode}
\usepackage{comment}

 \newcommand{\kPerp}{k_\perp}

\newcommand{\eps}{\varepsilon }

\renewcommand{\Im}{\frak{I} }

\renewcommand{\vec}[1]{\mathbf{#1}}

\newcommand{\tens}[1]{\mbox{\textsf{\textbf{#1}}}}
\newcommand{\dif}{\!\! \mathrm{d}}
\newcommand{\me}{\mathrm{e}}

\newcommand{\mi}{\mathrm{i}}

\renewcommand{\vec}[1]{\mathbf{#1}}

\newcommand{\Cadd}[1]{\textcolor{red}{#1}}

\begin{document}

\preprint{AIP/123-QED}
\newcommand{\acro}{Qerra} 

\title{Quantized Embedding Approaches for Collective Strong Coupling --- Connecting ab initio and macroscopic QED to Simple Models in Polaritonics} 

\author{Frieder Lindel}
  \email[Electronic address:\;]{frieder.lindel@uam.es}
\affiliation{Departamento de Física Teórica de la Materia Condensada, Universidad Autónoma de Madrid, E-28049 Madrid, Spain}
\affiliation{Physikalisches Institut, Albert-Ludwigs-Universit\"{a}t Freiburg, Hermann-Herder-Stra{\ss}e 3, D-79104, Freiburg, Germany}

  \author{Dominik Lentrodt}
  \affiliation{Physikalisches Institut, Albert-Ludwigs-Universit\"{a}t Freiburg, Hermann-Herder-Stra{\ss}e 3, D-79104, Freiburg, Germany}

  \author{Stefan Yoshi Buhmann}
  \affiliation{Institut f\"{u}r Physik, Universit\"{a}t Kassel, Heinrich-Plett-Stra{\ss}e 40, 34132 Kassel, Germany}

  \author{Christian Sch\"afer}
  \email[Electronic address:\;]{christian.schaefer.physics@gmail.com}
  \affiliation{Department of Physics, Chalmers University of Technology, 412 96 G\"oteborg, Sweden}

\begin{abstract}
Collective light-matter interactions have been used to control chemistry and energy transfer, yet accessible approaches that combine \textit{ab initio} methodology with large many-body quantum optical systems are missing due to the fast increase in computational cost for explicit simulations. We introduce an accessible \textit{ab initio} quantum embedding concept for many-body quantum optical systems that allows to treat the collective coupling of molecular many-body systems effectively in the spirit of macroscopic QED while keeping the rigor of \textit{ab initio} quantum chemistry for the molecular structure.
Our approach fully includes the quantum fluctuations of the polaritonic field and yet remains much simpler and more intuitive than complex embedding approaches such as dynamical mean-field theory. 
We illustrate the underlying assumptions by comparison to the Tavis--Cummings model.
The intuitive application of the quantized embedding approach and its transparent limitations offer a practical framework for the field of \textit{ab initio} polaritonic chemistry to describe collective effects in realistic molecular ensembles.
\end{abstract}

\maketitle

\section{Introduction}

May it be for catalysis, solar energy harvesting, or superconductivity, controlling the state of a material on demand and on the atomistic level defines a considerable area of modern science and technology.
In addition to the intrinsic electronic and phononic degrees of freedom of the material, their interplay with the electromagnetic environment can play a critical role.
Prominent examples include Floquet engineered materials~\cite{oka2019floquet,schafer2018insights,hubener2021engineering} and mode-selective chemistry~\cite{bloembergen1984energy}.
Coherently driving the material directly has two major downsides, (i) it requires energy, and (ii) it results in uncontrolled dissipation which tends to obscure the path towards the desired target state.
Electromagnetic resonators can reshape the continuous optical free-space spectrum into distinct optical modes, thus allowing a material inside to exchange energy only with specific modes.
At sufficiently strong interaction, the resonant modes of matter and cavity hybridize and result in polaritonic quasi-particle excitations.
Over the recent years, a plethora of changes in energy transfer~\cite{coles2014b, orgiu2015, zhong2016,herrera2016, fukushima2022inherent, wellnitz2022disorder,schafer2019modification, groenhof2019tracking, csehi2022competition, kumar2023extraordinary, timmer2023plasmon, engelhardt2023polariton} and chemical reactivity~\cite{Munkhbat2018, thomas2016, thomas2020ground, ahn_herrera_simpkins_2022, chen2022cavity, galego2016, fregoni2020strong, schafer2021shining} have been observed and established the field of polaritonic chemistry~\cite{genet2021inducing, garcia2021manipulating, simpkins2021mode}.

The strength of hybridization between light and matter scales approximately as $\sqrt{N_E/V}$, with the number of collectively coupled emitters $N_E$ and the effective quantization volume $V$ of the confined field. 
Strong coupling is reached if the hybridization energy exceeds all losses and the two bright polaritonic excitations can be clearly separated from their background.
Two prominent strategies exist to reach strong coupling: (i) minimizing $V$ by using e.g. subwavelength confined optical modes in (meta)plasmonic systems~\cite{chikkaraddy2016, wang2017coherent, Ojambati2019, arul2022giant}, and (ii) increasing the number of emitters $N_E$ that interact coherently with the confined resonator mode~\cite{Agranovich2003,coles2014b, orgiu2015, zhong2016,thomas2016,ahn_herrera_simpkins_2022}.
Polaritonic chemistry has primarily utilized the latter, which imposes, unfortunately, a high burden on the theory as the description of large ensembles becomes computationally overwhelming.
As a consequence, the literature is dominated by Tavis--Cummings-like models which replace the complex electronic structure with two levels and the resonator structure with a single harmonic oscillator.
Effective (embedding) extensions of such models have recently gained attention~\cite{10.1063/5.0002164,perez2022collective,10.1063/5.0101528} as they promise to compress complexity into intuitive expressions.
While simple quantum optical models have demonstrated remarkable success in describing the collective interaction, the goal to simultaneously capture changes in the chemical structure~\cite{schafer2021shining,schaefer24,simpkins2021mode,garcia2021manipulating} calls for new approaches that are able to combine sophisticated cavities~\cite{lalanne2018light} with the complex molecular restructuring during chemical reactions.
Importantly, chemistry is foremost local while collective interaction is intrinsically delocalized, i.e., a local reformulation would greatly ease the theoretical challenge.

In this spirit, quantizing macroscopic Maxwell's equations for all but a few molecules, or even a single molecule, is an attractive approach.
Macroscopic quantum electrodynamics (QED) is capable of describing a complex optical environment seen by a single molecule via the linear response functions of the surrounding molecules and the cavity structure. 
Screening effects are accounted for via local field corrections, the impact of which on collective strong coupling has not yet been explored.
Ref.~\citenum{schafer2022emb} introduced the Embedding radiation-reaction approach (Erra) which offers a simplified workflow compared to macroscopic QED and demonstrated that, within mean-field theory, the collective interaction can be approximately absorbed into an effective light-matter interaction that is trivially compatible with local \textit{ab initio} methods. 

In this manuscript, we clarify the potential and limitations of quantized embedding approaches based on macroscopic QED for collective strong coupling by connecting them to standard quantum optical models. 
To do so, we introduce in Sec.~\ref{sec:QEA} quantized embedding approaches and specifically in Sec.~\ref{sec:Qerra} the Quantized embedding radiation-reaction approach (\acro). Sec.~\ref{sec:ComparisonToTavis} demonstrates how \acro\ recovers the Tavis--Cummings Hamiltonian in a suitable limit as well as collective effects (superradiance). 
For simple cavity structures, we find that \acro\ offers a simple and straight-forward workflow (Sec.~\ref{sec:Workflow}) that provides the quantized description of a macroscopic number of realistic molecules interacting with the cavity in the low-excitation regime.
Our results in Sec.~\ref{sec:Extensions} demonstrate how embedding approaches offer a computationally feasible way to go beyond common approximations employed in standard quantum optics approaches to describe the local dynamic of a single molecule that is collectively strongly coupled to a realistic azopyrrole in chloroform solution. 
Sec.~\ref{sec:conclusion} concludes our discussion and provides an outlook over the remaining challenges and possible solution strategies for polaritonic embedding approaches.

\section{Quantized embedding approaches for collective strong coupling}
\label{sec:QEA}

We first give an overview of field quantization in general absorbing and dispersive media via macroscopic quantum electrodynamics (Section \ref{sec:FieldQuantization}). In Section~\ref{sec:EmbeddingApproaches}, we then demonstrate how this framework can be employed to examine collective strong coupling of a macroscopic number of complex emitters through embedding approaches. Section~\ref{sec:Workflow} and Fig.~\ref{fig:Workflow} presents a summary of the underlying workflow of these embedding approaches.

\subsection{Field quantization in absorbing environments}
\label{sec:FieldQuantization}

To find the electromagnetic field in the presence of linear macroscopic media\cite{scheel2009macroscopic,buhmann2013dispersionI}, a fruitful approach is offered by macroscopic electrodynamics. Here, the macroscopic number of matter degrees of freedom is encompassed into a dielectric function, namely the permittivity $\varepsilon(\vec{r},\omega)$\footnote{For simplicity, we assume throughout the manuscript that the permeability $\mu(\vec{r},\omega)$ of the media is given by $\mu(\vec{r},\omega) \approx 1$. The embedding approaches discussed here, however, can also be used to study magnetically responding media}. $\varepsilon$ relates the linear polarization field generated by the media to the external electric field. Canonically quantizing classical macroscopic electrodynamics \cite{philbin2010canonical} leads to the theoretical framework of macroscopic quantum electrodynamics \cite{scheel2009macroscopic,buhmann2013dispersionI}, which is capable of finding the quantized electromagnetic field in general absorbing and dispersive optical environments given by their permittivity.

The resulting quantized electric field operator reads\cite{scheel2009macroscopic,buhmann2013dispersionI}
\begin{align} \label{eq:E}
    \vec{E}(\vec{r}, \omega )  =  \mi \frac{\omega^2}{c^2}  \int \dif^3 r^\prime \sqrt{\frac{\hbar}{\pi \varepsilon_0} \mathrm{Im}[\varepsilon(\vec{r}^\prime , \omega)] } \tens{G}(\vec{r}, \vec{r}^\prime, \omega) \cdot \vec{f}(\vec{r}^\prime, \omega).
\end{align}
Here, $c$ is the speed of light in vacuum, $\varepsilon_0$ the vacuum permittivity, and $\vec{f}^{(\dagger)}(\vec{r}^\prime, \omega)$ are bosonic creation and annihilation operators satisfying 
\begin{align}
      [\vec{f} (\vec{r}, \omega), \vec{f} (\vec{r}^\prime, \omega^\prime)] & =   [\vec{f}^\dagger (\vec{r}, \omega), \vec{f}^\dagger (\vec{r}^\prime, \omega^\prime)] = \mathbf{0}, \\     [\vec{f} (\vec{r}, \omega), \vec{f}^\dagger(\vec{r}^\prime, \omega^\prime)] & = \boldsymbol{\delta}(\vec{r}- \vec{r}^\prime) \delta(\omega-\omega^\prime).
\end{align}
$\tens{G}$ is the dyadic Green tensor of the vector Helmholtz equation defined via
\begin{align}
     \left[  \nabla \times \nabla \times - \frac{\omega^2}{c^2}\varepsilon(\vec{r}, \omega) \right] \tens{G}(\vec{r}, \vec{r}^\prime, \omega) =    \boldsymbol{\delta}(\vec{r}-\vec{r}^\prime) , \label{eq:Green_def}
\end{align}
and the boundary condition $\tens{G}(\vec{r}, \vec{r}^\prime, \omega) \to 0 $ for \mbox{$|\vec{r}- \vec{r}^\prime |  \to \infty $}. The magnetic field operator can be found from Eq.~\eqref{eq:E} via $\vec{B} =\mi \nabla \times \vec{E}/ \omega $.

When the macroscopic QED field is coupled to a set of charged emitters at positions $\vec{r}_i$ and with dipole operators $\vec{d}_i$, the  corresponding light-matter Hamiltonian in the multipolar coupling scheme and the long-wave-length approximation reads\cite{buhmann2013dispersionI,feist2021macroscopic}
\begin{align}
    H = H_\mathrm{mat} + H_\mathrm{F} - \sum_{i}\vec{d}_i \cdot \vec{E}(\vec{r}_i).
\end{align}
Here, we have defined the matter Hamiltonian $H_\mathrm{mat}$ including the dipole self-energy term~\cite{schafer2019relevance} and $H_\mathrm{F}$ is the Hamiltonian of the medium-assisted field:
\begin{align}
      H_\mathrm{F} = \int \dif^3 r\int_0^\infty \dif \omega \,\hbar \omega \, \vec{f}^\dagger(\vec{r}, \omega) \cdot \vec{f}(\vec{r}, \omega).
\end{align}
To use macroscopic QED, one has to determine $\varepsilon(\vec{r},\omega)$ for the given setup and subsequently determine the Green tensor $\tens{G}$ by solving Eq.~\eqref{eq:Green_def}. $\tens{G}$ and $\varepsilon$ then fully determine the quantized electromagnetic field. For example, the spectral density of the field reads
\begin{align}\label{eq:SpectralDensity}
\tens{J}(\vec{r},\omega)
  =   \frac{\omega^2 }{\pi \varepsilon_0 c^2} \mathrm{Im}[\tens{G}(\vec{r}, \vec{r}, \omega)] \,,
\end{align}
which is known to determine the dynamics of a single quantum emitter coupled to the vacuum field at position $\vec{r}$.\cite{breuer2002theory} Beyond a single emitter, the dynamics of a few emitters (possibly strongly) coupled to the medium assisted field can analogously be found using macroscopic QED. However, finding the dynamics of macroscopically many emitters coupled to the four-dimensional continuum of field operators $\vec{f}^\dagger(\vec{r}, \omega)$ poses great computationally difficulties in general. Embedding approaches, as outlined in the following section, suggest a suitable path to compress the complexity of many-body systems to an effective single or few-particle problem.



\subsection{Ab initio embedding approaches}\label{sec:EmbeddingApproaches}




We consider the following setup, which is relevant, e.g., for polaritonic chemistry: A cloud of $N_\mathrm{E} \gg 1$ emitters (e.g. molecules) is placed inside a cavity. The basic idea of the embedding approaches is now to separate the system of $N_\mathrm{E} $ emitters and the cavity into microscopic components --- which are solved directly --- and macroscopic components --- which are described effectively. The microscopic part is the part of interest and in the following is considered to consist of just a single molecule (or a few) that can be described in detail by \textit{ab initio} methods. The macroscopic part consists of the remaining $N = N_\mathrm{E}-1$ molecules and the cavity. Both, the cavity mirrors and the remaining $N$ molecules, are assumed to dress the electromagnetic environment of the microscopic part via linear response theory, i.e., via an effective linear susceptibility $\boldsymbol{\chi}$.
Here, $\boldsymbol{\chi} =\boldsymbol{\chi}_\mathrm{mir}  +\boldsymbol{\chi}_\mathrm{mol}  $ consists of a part describing the cavity mirror ($\boldsymbol{\chi}_\mathrm{mir}$) and the molecules ($\boldsymbol{\chi}_\mathrm{mol}$), compare Fig.~\ref{fig:Workflow}. $\boldsymbol{\chi}_\mathrm{mol}$ can be obtained by calculating the polarizability $\boldsymbol{\alpha}$ of a single molecule via \textit{ab initio} methods and then employing a Clausius--Mossotti relation to find $\boldsymbol{\chi}_\mathrm{mol}$, e.g. in the dilute gas limit one finds\cite{jackson1999classical}
\begin{align} \label{eq:Clausius}
    \boldsymbol\chi_\mathrm{mol}(\vec{r}, \omega) = \frac{N}{V \varepsilon_0} \theta_V(\vec{r}) \boldsymbol\alpha(\omega).
\end{align}
Here, $V$ is the volume in which the emitters are located, and $\theta_V$ is the Heaviside theta function, i.e. $\theta_V(\vec{r}) = 1$ if $\vec{r} \in V$ and $\theta_V(\vec{r}) = 0$ if $\vec{r} \notin V$.

Having absorbed most of the emitters into the environmental susceptibility $\boldsymbol{\chi}_\mathrm{mol}$, we can use macroscopic QED as outlined in Section~\ref{sec:FieldQuantization} to quantize the effective electromagnetic environment $\tens{G}$ for the given $\boldsymbol{\chi}$, which now represents the cavity dressed by N emitters. What hampers the direct application of such an approach is the fact that the spectral density in Eq.~\eqref{eq:SpectralDensity} diverges in the coincidence limit (for $\vec{r}= \vec{r}^\prime$). Physically, this feature is caused by the coarse graining of the medium, which neglects that the field seen locally by the emitter differs from the macroscopic one due to the discreteness of the individual emitters\cite{onsager1936electric}. To account for these effects, local field corrections have to be employed, e.g. via the so-called real cavity model\cite{scheel1999spontaneous} as outlined in the next section.

\subsubsection{Full macroscopic QED embedding} \label{sec:FullEmbedding}

To account for local field effects, the real cavity model assumes that the space in the immediate vicinity of the emitter is empty, i.e.~that the emitter is placed in a small spherical cavity with radius $R_C$ centered around the position of the emitter inside the surrounding media, see Fig.~\ref{fig:Workflow} (left). The Green tensor of the new geometry, including the small cavity surrounding the emitter, can be determined in the limit that the radius $R_C$ satisfies $R_C k \ll 1$, where $k = \varepsilon(\omega) \omega/c$ is the wave-vector of the field. In this limit, one finds that the Green tensor inside the small cavity reads\cite{dung2006local}
\begin{align}
\begin{split} 
  \tens{G}(\vec{r},\vec{r}^\prime, \omega) &=   \tens{G}_\mathrm{vac}(\vec{r},\vec{r}^\prime, \omega)  +\tens{C}(\varepsilon, R_C, \omega) \\ \label{eq:LocalFieldCorrectedG}
  &+ \left( \frac{3 \varepsilon}{2\varepsilon + 1} \right)^2 \tens{G}^{(1)}(\vec{r}, \vec{r}^\prime, \omega),
\end{split}
\end{align}
where $\tens{G}_\mathrm{vac}$ is the free-space Green tensor, $\tens{G}^{(1)}$ is the scattering Green tensor of the surrounding medium without the small cavity around the emitter, and 
\begin{align}
    \tens{C}(\varepsilon,& R_C, \omega) =  \frac{k}{6\pi} \tens{1}\left\{  \frac{3 (\varepsilon-1)}{2\varepsilon + 1}  \frac{1}{(k R_C)^3} \right. \\
    &\left. + \frac{9 (\varepsilon-1)(4\varepsilon + 1)}{5(2\varepsilon + 1)^2}   \frac{1}{k R_C} + \mi \left[  \frac{9 \varepsilon^{5/2}}{(2\varepsilon + 1)^2} -1    \right] \right\}.\notag
\end{align}
The real cavity model has been used to study e.g.~corrections to the Purcell effect\cite{scheel1999spontaneous,tomavs2001local,dung2006local}, medium-assisted van-der Waals interactions\cite{Sambale2007}, and strong-coupling of a single emitter in cavity field with a background medium\cite{hien2011local}. However, it is not clear how to connect the full macroscopic QED embedding approach based on the real cavity model outlined here to standard quantum optics frameworks, such as the Tavis--Cummings Hamiltonian. Moreover, in the full macroscopic QED embedding approach, the scattering Green tensor of the cavity and the emitters $\tens{G}^{(1)}$ must be calculated for each type of emitter and each emitter density separately, as each new configuration of emitters inside the cavity results in slightly different mode-structures.
Both limitations can be lifted by introducing the streamlined embedding approach \acro. 

\subsubsection{Qerra: Good cavity, dilute emitter, and small volume approximation}\label{sec:Qerra}

Here, we follow Ref.~\citenum{schafer2022emb} and consider a scenario that closely resembles more idealized quantum optical setups. We assume that the molecules are confined to a region $V_\mathrm{mic}$ that is much smaller than the resonant wave-length of the cavity. We further consider the good cavity and dilute emitter limits in which one can safely ignore the diverging bulk contribution to the Green tensor as well as local field corrections, and approximate the bare Green tensor of the empty cavity $\overline{\tens{G}} \approx \overline{\tens{G}}^{(1)}$ by the scattering contribution $\overline{\tens{G}}^{(1)}$.
This neglects all direct interactions between the emitters that are not mediated by the cavity. In this limit, we can follow Ref.~\citenum{schafer2022emb} to obtain the Green tensor inside the small volume $V_\mathrm{mic}$ accounting for the surrounding $N$ emitters and the cavity via the following Lippmann-Schwinger equation\cite{buhmann2006born,buhmann2013dispersionII}
\begin{multline}\label{eq:GreensBornSeries}
\tens{G}(\textbf{r},\textbf{r}',\omega) =\overline{\tens{G}}(\textbf{r},\textbf{r}',\omega)\\
+ \frac{\omega^2}{c^2} \int_{V_\mathrm{mic}}dr'' \overline{\tens{G}}(\textbf{r},\textbf{r}'',\omega) \cdot \boldsymbol\chi_\mathrm{mol}(\textbf{r}'',\omega) \cdot \tens{G}(\textbf{r}'',\textbf{r}',\omega). 
\end{multline}
Here, $\vec{r},\vec{r}^\prime \in V_\mathrm{mic}$, and $\overline{\tens{G}}$ is the Green tensor of the empty cavity that we approximate by its scattering contribution, i.e., $ \overline{\tens{G}} \approx \overline{\tens{G}}^{(1)} $. Assuming that $\overline{\tens{G}} $ is approximately constant over the spatial extension of the cloud of molecules $V_\mathrm{mic}$, this equation can be solved via the dipole approximation, i.e., assuming that $\boldsymbol\chi_\mathrm{mol}(\textbf{r}'',\omega)  = \boldsymbol\chi_\mathrm{mol}(\omega) \delta (\vec{r^{\prime\prime}} )$, such that
\begin{multline}  \label{eq:BornDipole}
\tens{G}(\vec{r}, \vec{r}^\prime,\omega) \approx  \overline{\tens{G}}(\vec{r}, \vec{r}^\prime, \omega) \\ + V \frac{\omega^2}{c^2} \overline{\tens{G}}(\vec{r}, \vec{r}^\prime,\omega)\cdot \boldsymbol{\chi}_\mathrm{mol}(\omega) \cdot \tens{G}(\vec{r}, \vec{r}^\prime,\omega).
\end{multline}
This equation can be solved for $\tens{G}$ to yield 
\begin{align}  \label{eq:BornDipoleSolved}
\tens{G}(\vec{r}, \vec{r}^\prime,\omega) \approx& \left( [\overline{\tens{G}}(\vec{r}, \vec{r}^\prime, \omega)]^{-1}- V_\mathrm{mic} \frac{\omega^2}{c^2} \boldsymbol{\chi}_\mathrm{mol}(\omega)\right)^{-1}.
\end{align}
As discussed in Ref.~\citenum{schafer2022emb}, this approach is sufficient to describe the classical linear response of the entire system and provides a simple way to push \textit{ab intio} QED to macroscopically large values of $N$. This scheme can further be easily extended to not just a single type of molecule but, e.g., also the scenario where a molecule of interest is placed in a solvent. To this end, the polarizability would consist of a sum of the polarizabilities of the molecules of interest and the solvent molecule weighted by their relative density. Note that \acro~ provides solutions for any arbitrary number and mixture of emitters inside the resonator structure once the Green tensor of the bare resonator structure $\overline{\tens{G}}$ and the emitter polarizability $\boldsymbol{\alpha}$ have been determined. In contrast, in the full embedding approach accounting for local field corrections outlined in Section~\ref{sec:FullEmbedding}, the Green tensor of the cavity and the emitters must be calculated for every new concentration of the emitters.

So far, \acro's classical equivalent has been used to find the classical electromagnetic environment of a molecule inside a cloud of $N\gg 1$ molecules that is placed inside a cavity, given by a bare Green tensor consisting of just a single Lorentzian\cite{schafer2022emb}, which physically correspond to a perfectly single mode cavity. In the following, we use macroscopic QED to quantize the field in the embedding ansatz and show how this connects \acro~to standard quantum optics approaches in Section~\ref{sec:ComparisonToTavis}. Before doing so, we summarize the workflow of the macroscopic QED embedding approaches for collective strong coupling in the next section.

\subsection{Workflow of Quantized Embedding Approaches}
\label{sec:Workflow}

\begin{figure}
    \centering
    \includegraphics[width=1.\columnwidth]{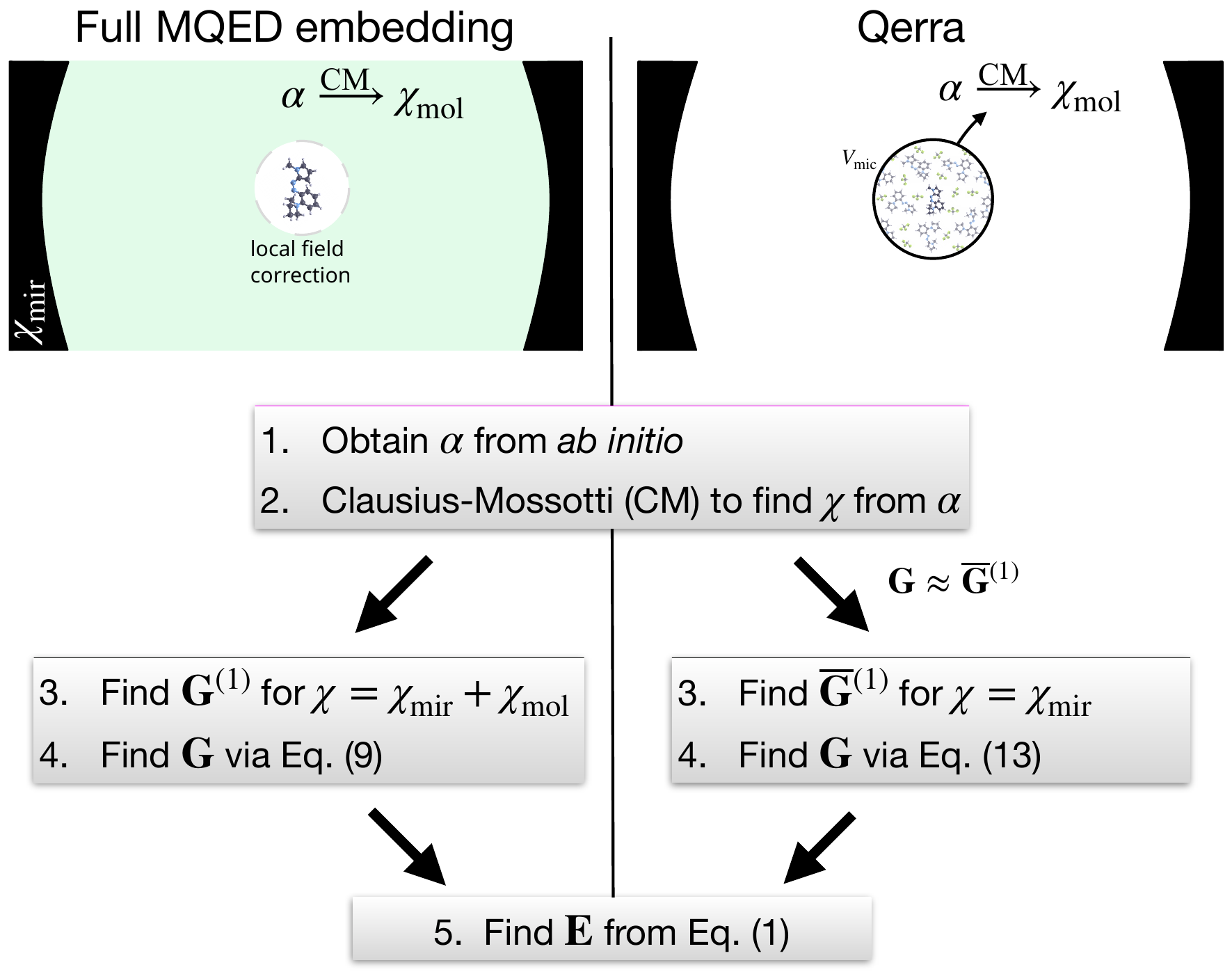}
    \caption{Workflow of embedding approaches for collective strong coupling based on macroscopic QED (MQED).}
    \label{fig:Workflow}
\end{figure}

The workflow of the embedding ansatz is summarized in Fig.~\ref{fig:Workflow}: In a first step, the polarizability of the individual molecules and the solvent is calculated in free-space from first principles using e.g.~time-dependent density-functional theory. 
The susceptibility $\boldsymbol{\chi}_\mathrm{mol}$ of the ensemble of $N$ molecules is then found for example via a Clausius--Mossotti type relation (compare Eq.~\eqref{eq:Clausius}). The Green tensor $\tens{G}(\vec{r}_{N_\mathrm{E}},\vec{r}_{N_\mathrm{E}},\omega) \equiv \tens{G}(\omega) $ of the optical environment of the singled-out emitter at position $\vec{r}_{N_\mathrm{E}}$ can be obtained either via the full macroscopic QED embedding approach (see Section~\ref{sec:FullEmbedding}) or via \acro~(see Section~\ref{sec:Qerra}). Once $\tens{G}(\omega) $ and the susceptibility of the environment are determined, the resulting quantized field is given by macroscopic QED via Eq.~\eqref{eq:E}. We have thus reduced the problem of a molecule interacting with a cavity of arbitrary geometry and material as well as a macroscopic number of other molecules to the interaction of a single molecule with a dressed multimode quantized field. 

The resulting multimode macroscopic QED field can be utilized directly in \textit{ab initio} calculations\cite{svendsen2021combining,svendsen2024ab}. Alternatively, one can further reduce the complexity of the four-dimensional continuum of field modes given by $\vec{f}(\vec{r},\omega)$ to only a few dominant modes of the field using a suitable few-mode model\cite{medina2021few,Lentrodt2020a,Menczel2024_arxiv,Lentrodt2023,Tamascelli2018,Lamprecht1999}. 

The embedding approaches outlined here allow one to include realistic cavity models (described through a general Green tensor), go beyond the rotating-wave approximation, and account for the full molecular structure provided from first principles, yet the computational complexity remains constant for any number of molecules $N$.

\section{Comparison to the Tavis--Cummings model} \label{sec:ComparisonToTavis}

Collective strong coupling is routinely studied via the Tavis--Cummings model and extensions thereof such as the Holstein--Tavis--Cummings model\cite{ribeiro2018polariton,fregoni2022theoretical}. In such approaches, the electronic degrees of freedom of the emitters are approximated by two-level systems that are all located at the same position inside the cavity, the cavity field is given by a lossless single harmonic mode.  

How does the embedding approach introduced in the last section compare to these standard quantum optics methods? To analyze this question, we consider $N$ two-level systems resonantly coupled to a single-mode cavity, and assume that the rotating-wave approximation applies. We consider only one polarization direction, such that we include only one diagonal element $\mathsf{G}$ of the Green tensor. For a single mode field, it reads \cite{lalanne2018light,Kristensen2020}
\begin{align} \label{eq:GreensSingleMode}
\overline{\mathsf{G}}(\omega) = \frac{f_1^2}{ \omega_c- \omega - \mi \gamma_c} \boldsymbol 1.
\end{align}
Here, $\omega_c$ is the frequency of the cavity, $\gamma_c$ its decay rate, and the constant $f_1$ determines the resulting light-matter coupling strength, see below. 

We can now follow different approaches: (i) start with a single mode approximation of the electric field resulting from Eq.~\eqref{eq:GreensSingleMode} to find the dynamics via the resulting Tavis--Cummings Hamiltonian; (ii) use the \acro~embedding workflow from above, i.e. find the new, polaritonic multimode field for the cavity coupled to $N$ molecules; (iii) we can use the Tavis--Cummings Hamiltonian as in (i), but additionally apply the Holstein-Primakoff approximation, which is valid in the low-excitation limit. 
These three approaches are outlined and contrasted in the subsequent three sections. 

 \subsection{Exact diagonalization of the Tavis--Cummings model}\label{sec:tav_exact}
 
Coupling $N_\mathrm{E}$ two-level systems with transition frequency $\omega_\mathrm{A} = \omega_c$ to the field given by the Green tensor in Eq.~\eqref{eq:GreensSingleMode} leads to the following Tavis--Cummings model\cite{buhmann2013dispersionII,medina2021few}
\begin{align}\label{eq:HTC}
    H_\mathrm{TC} = \omega_c a^\dagger a + \sum_{j=1}^{N_\mathrm{E}}\omega_A \sigma_z^{(j)} + g (a^\dagger \sigma^{(j)}_- + a \sigma_+^{(j)}),
\end{align}
where the coupling strength is given by 
\begin{align} \label{eq:gtof1}
g= f_1 d \omega_c /\sqrt{\varepsilon_0 \hbar} c
\end{align}
with $d$ the dipole moment of one of the two-level systems. Here, we have neglected dissipation of the cavity field by considering the limit $\gamma_c \to 0$. The Hamiltonian in Eq.~\eqref{eq:HTC} preserves the total number of excitations, such that $ H_\mathrm{TC} $ can be decomposed into subspaces with fixed number of excitations. In the following, we focus on the single-excitation subspace of $H_\mathrm{TC}$ given by $ H_\mathrm{TC}^{(1)} $. $H_\mathrm{TC}^{(1)}$ consists of: the single excited state $\ket{1}$ of the cavity; the bright state $\ket{B} \equiv N^{-1/2}\sum_i \ket{e_i}$, where $\ket{e_i}$ is the excited state of the $i$th emitter; the dark state manifold $\ket{D}$ that spans the subspace of the Hilbert space of the $N$ emitters with one excitation orthogonal to $\ket{B}$; and the excited state $\ket{e_{N_\mathrm{E}}}$ of the singled-out emitter. As the dark states $\ket{D}$ are not coupled to the other degrees of freedom, we neglect them in the following. As outlined in the supplemental material of Ref.~\citenum{schafer2022emb}, the resulting single-excitation Hamiltonian can be expressed in the basis $\{ (\ket{1} + \ket{B})/\sqrt{2} , (\ket{1} - \ket{B})/\sqrt{2}, \ket{e_{N_\mathrm{E}}} \}$ as
\begin{align} 
    H_\mathrm{TC}^{(1)} =\left(  \begin{array}{ccc}
    \omega_+^\mathrm{TC}     & 0  &  g/\sqrt{2}  \\
      0   & \omega_-^\mathrm{TC}   & g/\sqrt{2} \\
   g/\sqrt{2} & g/\sqrt{2} & \omega_{\mathrm{A}} \end{array} \right).
\end{align}
Here, $\omega_\pm^\mathrm{TC} = \omega_c \pm g \sqrt{N}$. We see that the problem reduces to two modes with frequencies $\omega_\pm^\mathrm{TC}$ coupled to the $N_E$th emitter with coupling strength $g/\sqrt{2}$. This is an exact solution of the model constrained to the single excitation manifold. 

 \subsection{Embedding approach} 

We next treat the same problem via the \acro~workflow (see the right hand side of Fig.~\ref{fig:Workflow}). The polarizability of a two-level system within the rotating-wave approximation is given by (we only consider one polarization direction) \cite{buhmann2013dispersionI}
\begin{align} \label{eq:PolTwoLevel}
 \alpha(\omega) = \frac{d^2}{\hbar} \lim_{\epsilon \to 0} \frac{1}{\omega_\mathrm{A} - \omega - \mi \epsilon}.
\end{align}
Using the Clausius--Mossotti relation in the dilute gas limit \eqref{eq:Clausius}, we find $\chi_\mathrm{mol}$ from the polarization in Eq.~\eqref{eq:PolTwoLevel}. Inserting $\chi_\mathrm{mol}$ and the single-mode Green tensor in Eq.~\eqref{eq:GreensSingleMode} into Eq.~\eqref{eq:BornDipoleSolved}, we find\footnote{Note, that in Eq.~\eqref{eq:GreensSingleMode4} the influence of the two-level systems (second term in the denominator) stays frequency dependent even for large detunings. This is an artifact of the rotating-wave approximation. The full polarizability reads $\alpha(\omega) = \frac{d^2 \omega_A }{\omega^2 - \omega_A^2+ \mathrm{i} \epsilon}$ such that 
\begin{align} \label{eq:GreensSingleMode3}
G(\omega) = \frac{1}{\frac{ \omega- \omega_c}{f_1^2} - \frac{N d^2 \omega^2 \omega_A}{c^2 \varepsilon_0  (\omega^2- \omega_A^2) }},
\end{align}
leading to a constant shift for the resonance frequency of the cavity for large detunings, i.e. $\omega \gg \omega_c$. This has a simple interpretation: Far from any resonances the two-level systems lead to a constant refractive index in the cavity, which shifts its resonances.}
\begin{align} \label{eq:GreensSingleMode4}
\mathsf{G}(\omega)  = \frac{1}{\frac{ \omega- \omega_c}{f_1^2} - \frac{N d^2 \omega^2}{c^2 \varepsilon_0 \hbar (\omega_\mathrm{A}- \omega) }}.
\end{align}
For $\omega_c = \omega_A$, this expression can be rewritten as
\begin{align}\label{eq:GThirdRow}
\mathsf{G}(\omega)  
& = \frac{f_1^2}{2} \left(  \frac{1}{1- \frac{\Omega_\mathrm{R}}{\omega_c}} \frac{1}{\omega - \omega_-} +  \frac{1}{1 + \frac{\Omega_\mathrm{R}}{\omega_c}} \frac{1}{\omega - \omega_+ } \right), 
\end{align}
with 
\begin{align}\label{eq:RabiFrequ}
\Omega_\mathrm{R}  &= \frac{f_1 \sqrt{N} d \omega_c}{c \sqrt{\varepsilon_0 \hbar } } = g \sqrt{N}, \\
\omega_\pm  & = \frac{\omega_c}{1 \pm \Omega_\mathrm{R}/\omega_c }.
\end{align}
We see that the resulting Green tensor in Eq.~\eqref{eq:GThirdRow} consists of two Lorentzian line shapes, which exactly correspond to two field modes with frequencies $\omega_\pm$. The coupling strengths $g_\pm$ of these two field modes to the singled-out emitter can be obtained from comparing Eq.~\eqref{eq:GThirdRow} with Eq.~\eqref{eq:GreensSingleMode} as well as using Eq.~\eqref{eq:gtof1}:
\begin{align}
      g_\pm  & = \frac{g}{\sqrt{2}} \frac{1}{\sqrt{1 \pm \frac{\Omega_\mathrm{R}}{\omega_c}}} .
\end{align}
We thus find that the dynamics of the singled-out emitter $N_\mathrm{E}$ coupled to the embedding environment, consisting of the cavity field mode and the $N$ other emitters, are governed by the effective Hamiltonian
 \begin{multline} \label{eq:HE}
    H_\mathrm{\acro}  =  \omega_{\mathrm{A}} \sigma^{(N_\mathrm{E})}_z  
    +  \sum_{i = \pm} \big[\omega_i a_i^\dagger a_i  + g_i (a \sigma^{(N_\mathrm{E})}_+ + a^\dagger \sigma^{(N_\mathrm{E})}_- )\big] .
\end{multline}
The single excitation manifold of this Hamiltonian reads
\begin{align}\label{eq:Ecoupling}   
   H_\mathrm{\acro}^{(1)}  &  = \left(  \begin{array}{ccc}
    \omega_+     & 0  &  g_+  \\
      0   & \omega_-   & g_- \\
   g_+ & g_- & \omega_{\mathrm{A}} \end{array}.  \right).
 \end{align}
Far from the ultra-strong coupling regime, we have $\frac{\Omega_\mathrm{R}}{\omega_c} \ll 1$, in which case we find
 \begin{align} 
    \omega_\pm & = \frac{\omega_c}{1 \pm \Omega_\mathrm{R}/\omega_c } \approx   \omega_c \pm  \Omega_\mathrm{R} = \omega_\pm^\mathrm{TC}  ,\\ \label{eq:gpm}
      g_\pm  & = \frac{g}{\sqrt{2}} \frac{1}{\sqrt{1 \pm \frac{\Omega_\mathrm{R}}{\omega_c}}} \approx \frac{g}{\sqrt{2}} \left(1 \mp \frac{\Omega_\mathrm{R}}{\omega_c}\right) \approx \frac{g}{\sqrt{2}}.
\end{align}
In this limit, we find 
\begin{align}
   H_\mathrm{\acro}^{(1)}  = H^{(1)}_\mathrm{TC}  .
\end{align}
This shows that for any number $N$ of the emitters, the dynamics restricted to the single-excitation manifold obtained via the embedding approach and via the exact diagonalization of the Tavis--Cummings Hamiltonian are exactly the same, as long as one does not enter the ultra-strong coupling regime ($\Omega_\mathrm{R}/\omega_c \ll 1$).
Note that the difference is here the renormalization of the transition, which is present in the embedding approach, but missing in the Tavis--Cummings Hamiltonian.

 \subsection{Low-excitation approximation of the Tavis--Cummings model}
 
 We further simplify the Tavis--Cummings Hamiltonian in Eq.~\eqref{eq:HTC} using the low-excitation approximation. To this end, we employ the Holstein--Primakoff transformation\cite{holstein1940field}. 
 
 We start by rewriting the Tavis--Cummings Hamiltonian in Eq.~\eqref{eq:HTC} using collective spin operators for the $N$ two-level systems, i.e. $J_z = \sum_{j=1}^{N} \sigma_z^{(j)}$, $J_{\pm} = \sum_{j=1}^{N} \sigma^{(j)}_\pm $ such that
\begin{multline}\label{eq:HTCFirst}
    H_\mathrm{TC} = \omega_c a^\dagger a  + \omega_A J_z + g(J_+ a + J_- a^\dagger) \\ +  \omega_A \sigma_z^{(N_\mathrm{E})} + g (a^\dagger \sigma^{(N_\mathrm{E})}_- + a \sigma_+^{(N_\mathrm{E})}).
\end{multline}
Expanding the $J$ operators in terms of bosonic operators $b$ via $J_z = b^\dagger b - N/2 \to  b^\dagger b  $ (in the last step we dropped a constant energy shift), $J_+ = b^\dagger \sqrt{N - b^\dagger b}  $, and $J_- =  \sqrt{N- b^\dagger b} b $ we find
\begin{multline}\label{eq:HTCSecond}
    H_\mathrm{TC} = \omega_c a^\dagger a  + \omega_A b^\dagger b + g(b^\dagger \sqrt{N - b^\dagger b} a  \\+  \sqrt{N - b^\dagger b} b a^\dagger)  +  \omega_A \sigma_z^{(N_\mathrm{E})} + g (a^\dagger \sigma^{(N_\mathrm{E})}_- + a \sigma_+^{(N_\mathrm{E})}).
\end{multline}
Note, that this is still an exact transformation of the Tavis--Cummings Hamiltonian. We next apply the low-excitation approximation assuming that only a negligible fraction of the atoms is excited, i.e. $b^\dagger b \ll N$ such that $\sqrt{N - b^\dagger b} \approx \sqrt{N}$. We find
\begin{multline}\label{eq:HHP}
    H_\mathrm{HP} = \omega_c a^\dagger a  + \omega_A b^\dagger b + g(b^\dagger \sqrt{N} a  \\+  \sqrt{N} b a^\dagger)  +  \omega_A \sigma_z^{(N_\mathrm{E})} + g (a^\dagger \sigma^{(N_\mathrm{E})}_- + a \sigma_+^{(N_\mathrm{E})}).
\end{multline}
 We can diagonalize the part of the Holstein--Primakoff Hamiltonian which describes the $N$ two-level systems and the bare cavity mode by introducing new polaritonic, bosonic annihilation operators $f_+ = (a + b)/\sqrt{2}$ and $f_- = (a - b)/\sqrt{2}$. We find
\begin{multline}\label{eq:HHPlikeE}
    H_\mathrm{HP} = \omega_A \sigma_z^{(N_\mathrm{E})} + \sum_{i=\pm} \bigg[ \omega_i^\mathrm{TC} f_i^\dagger f_i  \\
    + \frac{g}{\sqrt{2}} \left( f_i ^\dagger \sigma^{(N_\mathrm{E})}_- + f_i \sigma_+^{(N_\mathrm{E})} \right) \bigg].
\end{multline}
Note that the low-excitation approximation is exact for any number of atoms $N$ if one is restricted to the single-excitation manifold considered in Sec.~\ref{sec:tav_exact}, i.e.
\begin{align} 
    H_\mathrm{HP}^{(1)} =\left(  \begin{array}{ccc}
    \omega_+^\mathrm{TC}     & 0  &  g/\sqrt{2}  \\
      0   & \omega_-^\mathrm{TC}   & g/\sqrt{2} \\
   g/\sqrt{2} & g/\sqrt{2} & \omega_\mathrm{A} \end{array} \right) = H_\mathrm{TC}^{(1)}.
\end{align}
Furthermore, we find that 
\begin{align}
    H_\mathrm{\acro} = H_\mathrm{HP},
\end{align}
 holds in general, not only for the first excitation manifold. The embedding ansatz therefore leads to exactly the same model as the Tavis--Cummings model in the low-excitation approximation for any $N$ and any number of excitations, as long as $\Omega_\mathrm{R}/\omega_c \ll 1$. This correspondence shows that the embedding ansatz is valid whenever the low-excitation approximation is valid, i.e. for $b^\dagger b \ll N$. 
 Most experiments in polaritonic chemistry are performed in this low-excitation limit, many of them even in the "dark" for $N \gg 10^6$, such that \acro\ promises reasonably accurate predictions for as long as direct intermolecular interactions are of minor relevance.
 
Collectively coupled polaritonic systems at low pumping, i.e., in the low-excitation regime, represent therefore harmonic, i.e., classical, environments. \acro\ extends this statement from simple models to the realms of \textit{ab initio} QED by allowing for realistic response functions of the emitters and the cavity mirrors with only little additional computational cost. The harmonic nature of the environment validates also the usage of macroscopic QED, as the latter is based on the linear optical response.

\subsection{Collectivity} \label{secsub:Collective}

We have used \acro\ to single out one molecule acting as "impurity" which is embedded in the environment dressed by all other molecules via linear response theory. What is unclear so far is how much collectivity remains by following this procedure.

In case of $N_\mathrm{E}$ two-level systems, superradiance occurs when we prepare the two-level systems in a coherent superposition state $\ket{+} =(1/\sqrt{N_\mathrm{E}})\sum_{i = 1}^{N_\mathrm{E}} \sigma_+^{i}|\{0\} \rangle =(1/\sqrt{N_\mathrm{E}}) (J_+ + \sigma_+^{(N_\mathrm{E})} )|\{0\} \rangle $. When calculating the transition rate from $\ket{+,0} \to \ket{0, 1} $ and from $\ket{1_j,0} \to \ket{0, 1} $, where $\ket{1_j}$ is the state where atom $j$ is in the excited state and all other atoms are in the ground state and $\ket{0}$ and $\ket{1}$ correspond to the zero and one-photon states with respect to $a$, we find the transition matrix elements
\begin{align}
    \frac{\braket{0,1 | H_\mathrm{TC}|+,0} }{\braket{0,1 | H_\mathrm{TC}|1_j,0}}= \sqrt{N_\mathrm{E}}. 
\end{align}
The enhancement of $\sqrt{N_\mathrm{E}}$ compared to the single atom decay rate is commonly known as superradiance. We immediately see from the Holstein--Primakoff Hamiltonian $H_\mathrm{HP}$ in Eq.~\eqref{eq:HHP} that $\ket{+}$ couples with coupling strength $g \sqrt{N}$ to the single field mode. Since the Holstein--Primakoff Hamiltonian is equivalent to the one from our embedding approach, it immediately follows that collective effects such as superradiance are also included in the latter approach. To show this explicitly, we have to express $\ket{+}$ in terms of the new eigenmodes defined by $f_i^{(\dagger)}$ and $\sigma_\pm^{(N_\mathrm{E})}$. We find in the low-excitation limit 
\begin{align}
    \ket{+} 
    & = \frac{1}{\sqrt{N_\mathrm{E}}} \left[\sqrt{\frac{N}{2}} (f_+^\dagger - f_-^\dagger )   +  \sigma_+^{(N_\mathrm{E})} \right] \ket{\{0\}}.
\end{align}
Calculating the transition rate using the embedding Hamiltonian $H_\mathrm{\acro}$ in Eq.~\eqref{eq:HE} [which it can be connected directly to the Holstein--Primakoff Hamiltonian $H_\mathrm{HP}$ in its form \eqref{eq:HHPlikeE}], we get the same enhancement of $\sqrt{N_\mathrm{E}}$: The single photon state can be expressed as $\ket{1,0} =(f^\dagger_+ + f^\dagger_-)/\sqrt{2} \ket{0,0}$ and we find 
\begin{align}
    &\braket{0,1| H_\mathrm{\acro} |+,0} = \notag\\
    &~~\braket{0,0|\sqrt{ \frac{N}{4 N_\mathrm{E}}} (f_+ + f_-) H_\mathrm{\acro} (f_+^\dagger - f_-^\dagger)  |0,0} \notag \\
    &~~+ \braket{0,0|\sqrt{ \frac{1}{2 N_\mathrm{E}}} (f_+ + f_-)H_\mathrm{\acro} \sigma_+^{(N_\mathrm{E})}  |0,0}  \approx g \sqrt{N_\mathrm{E}}, \notag
\end{align}
where the last equality sign holds for $\omega_\pm = \omega_c \pm \Omega_\mathrm{R}$ with $\Omega_\mathrm{R} = g \sqrt{N}$, compare Eq.~\eqref{eq:RabiFrequ}. This conclusion is similar to Ref.~\citenum{10.1063/5.0002164} and linear dispersion theory \cite{Born1980}, which has been shown to cover strong coupling \cite{Zhu1990} as well as collective effects.

\section{Applications} \label{sec:Extensions}

Having established a connection between the \acro~embedding approach and the Tavis--Cummings Hamiltonian, we next illustrate how \acro~can be used to go beyond the limitations of the Tavis--Cummings model with only little extra computational effort. First, in Section~\ref{sec:GoodCavityLimit}, we show how \acro\ can be used to account for detuning of the two-level emitters and emitter losses. Then, we use \acro\ to study collective strong coupling of a more realistic molecular system in Section~\ref{sec:RealsiticMolecular}. In Section~\ref{sec:EvenlyFilled}, we apply the full macroscopic QED embedding approach (Section~\ref{sec:FullEmbedding}) to a setup in which the emitters are not localized in the center of the cavity, but are evenly distributed between the mirrors of a planar Fabry--P\'erot resonator. We account for dispersion and absorption in the gold mirrors of the cavity and further include local field corrections. For all applications, we only consider a single polarization direction of the field.


\begin{figure*}
    \centering
    \includegraphics[width=1.\linewidth]{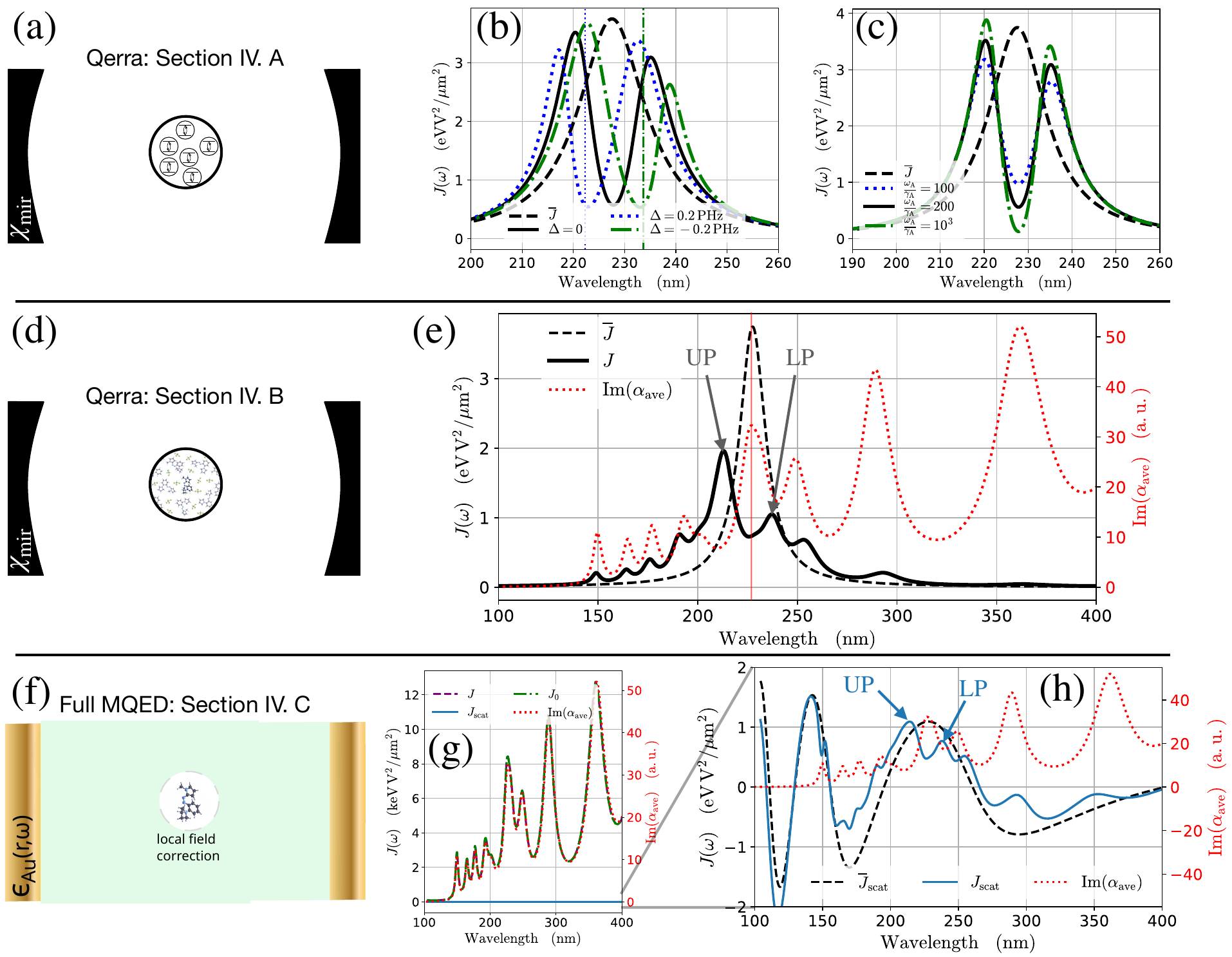}
    \caption{\textit{Application of embedding approach for collective strong coupling.} (a--c) Collective strong coupling of $N = 6\times 10^6$ two level systems with dipole moment $d = 2.25\,$D to a single mode cavity. (a) Scheme of the setup; (b,c) We used the \acro\ workflow (see Section \ref{sec:Qerra}) to find the spectral density $J$ of the optical environment of the singled-out emitter. $\overline{J}$ is the spectral density of the empty cavity given in terms of the Green tensor in Eq.~\eqref{eq:GreensSingleMode} with frequency $\hbar \omega_\mathrm{c} = 5.44\,$eV (corresponding to $227.8$\,nm), quality factor $ Q= \omega_\mathrm{c}/\gamma_c = 25.8$, and we set $f_1$ such that the collective Rabi splitting $\Omega_R /\omega_c = \sqrt{N}g/\omega_c = 0.13$. $J$ is obtained via Eq.~\eqref{eq:BornDipoleSolved} for (a) different detunings $\Delta = \omega_c-\omega_A$ and $\omega_A/\gamma_A = 200$, and (b) for fixed $\Delta =0$ and different decay rates $\gamma_A$. (d,e) The same single mode cavity as in (a--c) is coupled to ortho-azopyrrole in trans- and cis-configuration (15\% and 5\% of the total number of coupled molecules $N = 10^8$) in chloroform solution (80\% of coupled molecules), with polarizability $\alpha_\mathrm{ave}$ shown in (e) by the red dotted line. In (e), the dressed spectral density $J$ shows a splitting of the resonances of the empty cavity (black dashed) and the azopyrrole molecules at $227.8$\,nm (indicated by the vertical red line) into upper (UP) and lower (LP) polaritons. (f--h) Full macroscopic QED embedding approach (see Section~\ref{sec:FullEmbedding}). (f) Setup under consideration; (g) the full dressed spectral density $J$, its scattering ($J_\mathrm{sc}$) and bulk ($J_0$) contribution, and the polarizability of the azopyrrole in chloroform solution [same as in (e)]. (h) Zoom in on the scattering part of the dressed spectral density. Additionally, we show the spectral density of the empty cavity to illustrate the splitting of the cavity and emitter resonances at $227.8$\,nm into UP and LP. In (g,h), the Fabry--P\'erot cavity has length $L = 388\,$nm, the permittivity of the gold mirrors is given in Eq.~\eqref{eq:epsAu}, we use a number density of the emitters $\eta = 3\,\mathrm{nm}^{-3}$, and set the radius of the local field correction cavity to $R_C = 1\,$nm.}
    \label{fig:Application}
\end{figure*}

\subsection{\acro: two-level emitters} \label{sec:GoodCavityLimit}

We use the \acro\ workflow (see right hand side of Fig.~\ref{fig:Workflow}), to study collective strong coupling of two-level emitters to a single cavity mode, i.e.~$\overline{\mathsf{G}}$ is given by Eq.~\eqref{eq:GreensSingleMode}. To this end, we show in Fig.~\ref{fig:Application} (b,c) the spectral density \eqref{eq:SpectralDensity} to which the singled-out emitter couples, consisting of the single-mode cavity with frequency $\omega_c$ and decay rate $\gamma_c$, and the $N$ two-level emitters with frequency $\omega_A$, decay $\gamma_A$, and polarizability
\begin{align} \label{eq:PolTwoLevel2}
 \alpha(\omega) = \frac{d^2}{\hbar}  \frac{1}{\omega_\mathrm{A} - \omega - \mi \gamma_A}.
\end{align}
Note that the decay rate $\gamma_A$ (Purcell effect) and the frequency $\omega_A$ (environment-induced frequency shift) are affected by the cavity and $N_E$ emitters via the embedding approach.
We see in Fig.~\ref{fig:Application}(a) that for zero detuning $\Delta \equiv \omega_c- \omega_A = 0$, the single Lorentzian of the cavity (dashed black line) splits into two polaritonic peaks separated by the Rabi frequency $\Omega_R$. The slight difference of the height of the upper and lower polariton indicates that in general the coupling strength of the singled-out emitter to the upper and lower polariton is not the same, i.e. $g_+ \neq g_-$, compare Eq.~\eqref{eq:gpm}. If the two-level systems are red or blue shifted from the cavity frequency, we find that the resonance frequencies of the polaritons are also red or blue shifted, respectively. Furthermore, the relative height of the two polaritonic peaks changes, meaning that a singled-out emitter in a blue-shifted (red-shifted) ensemble will couple more strongly to the lower polariton (upper polariton) [see blue and green line in Fig.~\ref{fig:Application}(a)].

When considering the impact of the decay rate $\gamma_A$ of the two-level systems onto the spectral density, we find that with decreasing decay rate of the emitter the two polaritonic resonances in the spectral density become narrower. This shows that with decreasing decay rate of the emitters, also the decay rate of the polaritonic modes decreases, as expected.

\subsection{\acro: realistic molecular example using \acro} \label{sec:RealsiticMolecular}


Having established \acro\ as a tool to study the collective strong coupling of two-level systems to a single-mode cavity, we next show how it can be used to analyze collective strong coupling of realistic molecular ensembles. To this end, we have calculated the polarizability of ortho-azopyrrole in trans- and cis-configuration (15\% and 5\% of the total number of coupled molecules) in chloroform solution (80\% of coupled molecules) using time-dependent density-functional theory in combination with an implicit polarization model for the chloroform solution, see Appendix~\ref{app:Polarizability} for details.
The isotropic average $\mathrm{Im}[\alpha_\mathrm{ave}(\omega)]$ is shown in Fig.~\ref{fig:Application} (e).

The cavity consists of a single linearly polarized cavity mode that is tuned in resonance with one of the resonances of the azopyrrole solution at $227.8\,$nm, compare black dashed line in Fig.~\ref{fig:Application} (e). The resulting spectral density of the \acro\ workflow is shown as black line in the same figure. We can clearly see that the single resonance of the cavity splits into a lower and upper polariton. As before, the lower polariton has a lower height resulting in a lower coupling to the singled-out emitter. The line shape of these two polaritonic modes differs from simple Lorentzians, indicating the necessity of multiple modes to fully describe them in an effective few-mode model \cite{medina2021few}. On top of these two polaritonic resonances, we find that the dressed spectral density also contains additional features corresponding to the other resonances of the azopyrrole solution in the vicinity of the cavity frequency. As all of these features only consist of a single peak per resonance, we find that none of the other resonances couple collectively strongly to the cavity mode (the decay rate is bigger than the collective coupling strength, resulting in no visible Rabi splitting).


\subsection{Full macroscopic QED embedding with realistic cavity and molecular system} \label{sec:EvenlyFilled}

As a last example, we consider the setup illustrated in Fig.~\ref{fig:Application} (f). Here, the azopyrrole solution, already considered in the last subsection, fills all of the space between the two mirrors with permittivity $\eps_\mathrm{Au}$ of a Fabry--P\'erot cavity. The singled-out emitter is assumed to be located in the center between the two mirrors. To treat this problem, we make use of the full macroscopic QED embedding approach including local field corrections, see Section~\ref{sec:FullEmbedding}. We show the resulting spectral density for the singled-out emitter in Figs.~\ref{fig:Application}(g,f), see Appendix \ref{app:FullMQEDe} for numerical details. 
Local field corrections dominate the spectral density in this system, suggesting that Coulombic polarization-corrections routinely used in quantum chemistry for the description of solute-solvent dynamics (also used in Sec.~\ref{sec:RealsiticMolecular}) provide a major contribution to the dynamic of the impurity molecule.
In Fig.~\ref{fig:Application}(g), we further distinguish the scattering part of the spectral density, given by\footnote{We only consider a single polarization direction in-plane with the cavity mirrors for simplicity.}
\begin{align} \label{eq:Jsc}
    J_\mathrm{sc}(\vec{r},\omega)
  =   \frac{\omega^2 }{\pi \varepsilon_0 c^2} \mathrm{Im}\left[\left( \frac{3 \varepsilon}{2\varepsilon + 1} \right)^2 \mathsf{G}^{(1)}(\vec{r}, \vec{r}, \omega)\right],
\end{align}
from the bulk part $ J_0(\vec{r},\omega) = J(\vec{r},\omega) - J_\mathrm{sc}$, compare Eqs.~\eqref{eq:LocalFieldCorrectedG} and \eqref{eq:SpectralDensity}. Note that the bulk part $J_0$ gives the spectral density in the absence of the cavity, such that $J_\mathrm{sc}$ represents all changes in the optical environment of the singled-out molecule induced by the cavity. We see that $J_\mathrm{sc} \ll J_0$ indicating that the single emitter 'feels' the cavity only to a small extend and its dynamics are instead dominated by its direct chemical and free-space environment.
This is no surprise, as for Tavis--Cummings-like models of collectively coupled ensembles each individual molecule contributes only to a minor extend and is also affected only to a minor extend. In such simple models, the effective coupling strengths acting on a single molecule has an upper bound defined by the bare cavity coupling, as is apparent from Eq.~\eqref{eq:Ecoupling} and following (see also Ref.~\citenum{schafer2022emb}). Under which conditions the cavity is able to induce local changes remains a matter of active debate.~\cite{schafer2022emb,Castagnola24,sidler2021perspective,mandal2023theoretical,schnappinger2023cavity}

Still, as shown in Fig.~\ref{fig:Application}(h), the local-field corrected scattering part of the spectral density $J_\mathrm{sc}$ \eqref{eq:Jsc} shows the splitting of the cavity resonance at $227.8$\,nm into a lower and upper polariton. Furthermore, as the cavity resonances of the Fabry--P\'erot cavity are very broad, also the influence of other emitter resonances onto $J_\mathrm{sc}$ are clearly visible.

\section{Conclusions} \label{sec:conclusion}
A comprehensive description of collective strong coupling from first principles is a defining challenge in the theoretical description of polaritonics.
Established models require a strong simplification of the molecular structure and even then become prohibitively expensive to solve numerically for increasing numbers of emitters.

Here, we discussed quantized embedding approaches for collective strong coupling which are based on macroscopic QED and in which the ensemble of emitters located in the cavity are split into an "impurity" and "environment" subsystem.
All environmental degrees of freedom, including their (in)homogeneous broadening, are then used to dress the electromagnetic mode-structure encoded in the dyadic Green tensor $\tens{G}[\chi]$.
Usage of the entire dyadic results in an embedding version of full macroscopic QED but necessitates local field corrections and the consistent treatment of free-space, direct longitudinal interactions between the emitters, and cavity induced interactions, which complicates its direct application.
If the scattering contribution to $\tens{G}$ dominates the mode-structure, i.e., working with high-Q cavities, and direct (mostly longitudinal) interactions can be ignored in the electromagnetic treatment (dilute limit), we can derive the simplified description \acro .
\acro\ is easy to combine with \textit{ab initio} quantum chemistry, and offers therefore an intuitive approach to transition into the realm of \textit{ab initio} QED.
Especially many of the recently developed quantum chemistry~\cite{haugland2020coupled,10.1063/5.0216993,flick2018light,schafer2021making} and non-adiabatic dynamic~\cite{groenhof2019tracking,doi:10.1021/acs.jctc.2c01154,hoffmann2019benchmarking,vendrell2018coherent,10.1063/5.0033338} methodologies could be pushed closer to experimental reality by (partially) invoking quantum embedding approaches.

Following macroscopic QED allows then to quantize the dressed environment and couple it to the remaining impurity subsystem that is solved in full microscopic rigour.
The impurity is coupled effectively to a set of quantized effective oscillator modes that represent the polaritonic eigenstates of the environment.
We further illustrated this by demonstrating that \acro\ is identical to the Tavis--Cummings model in the low-excitation limit and that collective (superradiant) effects are included in the single-excitation space.
However, quantum embedding approaches for collective strong coupling can be readily extended to more complex environments, such as a cavity filled with a solute-solvent mixture that overlaps spectrally.
We illustrated how such a complex environment affects the spectral density and how lossy cavities and emitters can be described with the full macroscopic QED embedding approach.
Future work should focus on exploring possibilities to account for longitudinal inter-molecular interactions, thus moving \acro\ closer to full macroscopic QED embedding, anharmonic corrections to the dynamic of the embedded ensemble, and full self-consistency between impurity and environment.

Quantized embedding approaches promise a theoretically rigorous description from first principles for large ensembles of molecules collectively coupled to electromagnetic resonators -- paving a way to interweave the intuitive description in quantum optical models with the complexity of \textit{ab initio} QED in a local framework suitable for QED chemistry.

\begin{acknowledgments}
We thank Andreas Buchleitner and Edoardo Carnio for insightful discussions. This work was funded by the Spanish Ministry for Science and Innovation-Agencia Estatal de Investigación (AEI) through the grant EUR2023-143478. D.L. gratefully acknowledges financial support from the Georg H. Endress Foundation and the DFG funded Research Training Group ``Dynamics of Controlled Atomic and Molecular Systems'' (RTG 2717). C.S. acknowledges funding from the Horizon Europe research and innovation program of the European Union under the Marie Sk{\l}odowska-Curie grant agreement no.\ 101065117.

Partially funded by the European Union.
Views and opinions expressed are, however, those of the author(s) only and do not necessarily reflect those of the European Union or REA.
Neither the European Union nor the granting authority can be held responsible for them.
\end{acknowledgments}

\section*{Author declarations}
\subsection*{Conflict of Interest}
The authors have no conflicts to disclose.

\subsection*{Author Contributions}
F. Lindel: Investigation (equal); Methodology (equal); Writing – original draft (equal). Writing – review \& editing (equal).
D. Lentrodt: Investigation (equal); Methodology (equal); Writing – original draft (equal). Writing – review \& editing (equal). 
S. Y. Buhmann: Supervision (equal); Writing – review \& editing (equal).
C. Sch\"afer: Conceptualization (equal); Investigation (equal); Methodology (equal); Supervision (equal); Writing – original draft (equal). Writing – review \& editing (equal).

\appendix

\section{Numerical Details} 

\subsection{Green tensor of a gold Fabry--P\'erot cavity filled with azopyrrole solution} \label{app:FullMQEDe}

To apply the full macroscopic QED embedding workflow to the setup shown in Fig.~\ref{fig:Application}(f), we use that the imaginary part of the free-space Green tensor in the coincidence limit is given by 
\begin{align}
 \mathrm{Im}[\tens{G}(\vec{r},\vec{r}, \omega)]   = \frac{\omega}{6\pi c} \boldsymbol{1}.
\end{align}
Moreover, we have to evaluate the scattering Green tensor of the Fabry--P\'erot cavity $\tens{G}^{(1)}$ completely filled with a azopyrrole solution with permittivity $\varepsilon_\mathrm{as} = 1 + \chi_\mathrm{mol}$, compare Eq.~\eqref{eq:LocalFieldCorrectedG}. We here only consider one polarization component $\mathsf{G}^{(1)}_{xx}$ of the field that is in-plane with the cavity mirrors. $\mathsf{G}^{(1)}_{xx}(\vec{r},\vec{r}^\prime, \omega)$ in the coincidence limit in the center of the cavity is given by \cite{buhmann2013dispersionI}
\begin{multline} \label{eqappG:GreenScatteringCavity}
\mathsf{G}_{xx}^{(1)}(\omega) = \frac{\mi}{4\pi^2} \int \dif^2 k_{\parallel} \frac{1}{k_\perp}  \\
 \times  \left[  \me^{2\mi \kPerp L} \left( \frac{r_s^2 }{D_s} \frac{k_y^2}{k^2} + \frac{r_p^2 }{D_p} \frac{k_z^2 k_x^2}{k^2 k_\parallel^2}  \right) \right. \\
 \left.  + \me^{\mi \kPerp L}  \left(  \frac{r_s}{D_s} \frac{k_y^2}{k^2}   - \frac{r_p}{D_p} \frac{k_z^2 k_x^2}{k^2 k_\parallel^2}   \right)          \right].
\end{multline}
Here, $L$ is the distance between the two mirrors, $k = \sqrt{\varepsilon_\mathrm{as}} \omega/c$, $k_\perp = \sqrt{k^2-k_\parallel^2}$ with $\mathrm{Im}[k_\perp] > 0$, $D_\sigma = 1- r_\sigma^2 \me^{2\mi k_\perp L}$ account for multiple reflections between the mirrors, and $r_p$ and $r_s$ are Fresnel reflection coefficients that read
\begin{align}
    r_p & = \frac{k_\perp - k_\perp^{(\mathrm{Au})}}{k_\perp + k_\perp^{(\mathrm{Au})}} ,\\
    r_s & = \frac{ \varepsilon_\mathrm{Au}(\omega) k_\perp - \varepsilon_\mathrm{as}(\omega) k_\perp^{(\mathrm{Au})}}{ \varepsilon_\mathrm{Au}(\omega) k_\perp + \varepsilon_\mathrm{as}(\omega) k_\perp^{(\mathrm{Au})}},
\end{align}
with $k_\perp^{(\mathrm{Au})} = \sqrt{\epsilon_\mathrm{Au} (\omega^2/c^2) - k_\parallel^2}$. The permittivity of the gold mirror $\varepsilon_\mathrm{Au}$ is given by a Drude model 
\begin{align} \label{eq:epsAu}
    \varepsilon_\mathrm{Au}(\omega) = 1 + \frac{\omega_p}{\omega^2 + \mi \omega \gamma_\mathrm{Au}},
\end{align}
with $\omega_p = 2.067\times 2\pi$\,PHz and $\gamma_\mathrm{Au} = 4.4491 \times 2\pi$\,THz \cite{blaber2009search}. We further introduce polar coordinates $k_x = k_\parallel \mathrm{sin}(\phi)$ and $k_y = k_\parallel \mathrm{cos}(\phi)$ and perform the resulting $\phi$ integral analytically. The remaining $k_\parallel$ integral has been evaluated numerically.

\subsection{Polarizability for azopyrrole solution} \label{app:Polarizability}
All polarizabilities have been calculated using Casida time-dependent density-functional theory with the ORCA5.0 code~\cite{neese2022software}.
Structures for trans-azopyrrole, cis-azopyrrole, and chloroform have been relaxed using the CAM-B3LYP functional~\cite{yanai2004new} in a def2-TZVPD basis including the implicit CPCM solvation model for chloroform~\cite{doi:10.1021/jp000156l}.

The 20 first roots of the Casida equations under Tamm-Dancoff approximation are then used to construct the polarizability according to\cite{ullrich2011}
\begin{align*}
\alpha_{ij}(\hbar\omega) = \sum_{k=1}^{20} \frac{2 \hbar \omega_{k0} d_{k0}^{i} d_{k0}^{j}}{(\hbar \omega + i \eta)^2 - (\hbar\omega_{k0})^2}
\end{align*}
with the transition dipoles $\textbf{d}_{k0}$ and transition energy $\hbar\omega_{k0}$ between ground state and S-state $k$. We include an artificial broadening of $\eta = 5\cdot 10^{-3}$ Hartree.

\bibliography{library}

\end{document}